\documentclass[preprint,showpacs,preprintnumbers,amsmath,amssymb,showkeys]{revtex4}
\usepackage{graphicx}% Include figure files
\usepackage{dcolumn}% Align table columns on decimal point
\usepackage{bm}% bold math

\begin{document}
\renewcommand{\thesection}{\arabic{section}}
\renewcommand{\thetable}{\arabic{table}}
\renewcommand{\thetable}{\Roman{table}}

\makeatletter
\renewcommand{\fnum@table}[1]{Table~\thetable. } % for tables if req.
\renewcommand{\fnum@figure}[1]{FIG~\thefigure. }
\makeatother

%\bibliographystyle{unsrt}
%\bibliographystyle{nature}
%\bibliographystyle{alpha}
%\title{\center{Electron wave functions on  $T^2$ in a static
%magnetic field of arbitrary direction}}

\bibliographystyle{apsrev}
\title{Coupling curvature to a uniform magnetic
field;\\ an analytic and numerical study }

\author{M. Encinosa }
%\affiliation{ The National High Magnetic Field Laboratory\\T-Physics Division \\
%205 Jones Hall \\ Tallahassee FL 32307}
\email{encinosa@cennas.nhmfl.gov}
 %\altaffiliation[Also at ]{Physics Department, XYZ University.}
%\author{Lonnie Mott}%
\affiliation{ Florida A\&M University Department of Physics \\
Tallahassee FL 32307} %\affiliation{ DRAFT}

\begin{abstract}
\setlength{\baselineskip}{14.0pt}
 The Schrodinger equation for an
electron near an azimuthally symmetric curved surface $\Sigma$ in
the presence of an arbitrary uniform magnetic field $\mathbf B$ is
developed. A thin layer quantization procedure is implemented to
bring the electron onto $\Sigma$, leading to the well known
geometric potential $V_C \propto h^2-k$ and a second
 potential that couples $A_N$, the component of $\mathbf A$
 normal to $\Sigma$  to  mean surface curvature, as well as a
 term dependent on the normal derivative of
 $A_N$ evaluated on $\Sigma$. Numerical
results in the form of ground state energies as a function of the
applied field in several orientations are presented for a toroidal
model.
\end{abstract}

\pacs{03.65Ge, 73.22.Dj}% PACS, the Physics and Astronomy
\keywords{torus, magnetic field, geometric potential}%Use showkeys class option if keyword
\maketitle
\setlength{\baselineskip}{12.0pt}

\section{Introduction}

Nanostructures with novel geometries have become the subject of a
large body of  experimental and theoretical work \cite{bellucci,
bulaev,chou, datta, encT2mag, filikhin, fuhrer, goker,gravesen,
gridin, gylfad, ivanov, latge, latil,  lorke1, lorke2, mendach,
pershin, qu, sasaki, sano, tempere,zhang}. Many of the fabricated
structures exhibit curvature on the nanoscale making once purely
theoretical investigations of quantum mechanics on curved and
reduced dimensionality surfaces relevant to device modelling.
Because magnetic field effects prove important to Aharanov-Bohm
and transport phenomena, the interplay of an applied field with
the curved regions of a nanostructure \cite{bellucci, chryssom,
datta,  encT2mag, goker, ivanov, latil, sasaki, sasaki2}
(particularly if the structure has holes) may be critical to a
complete understanding of the physics of a nanodevice element.

In a previous work \cite{encT2mag}  the Schrodinger equation for
an electron on a toroidal surface $T^2$ in an arbitrary static
magnetic field was developed and numerical results obtained. There
however, the electron was restricted ab-initio to motion on $T^2$,
which precluded the appearance of the well known geometric
potential $V_C \propto  (h^2 - k)$,
\cite{burgsjens,jenskoppe,dacosta1,dacosta2,matsutani,matsutani2,
duclosexner,bindscatt,popov, ouyang, midgwang,ee1,ee2, goldjaffe,
exnerseba,schujaff,clarbrac}  with $h,k$ the mean and Gaussian
curvatures, that arises via thin-layer quantization
\cite{golovnev}. Furthermore, if the degree of freedom normal
 to  $\Sigma$ (labelled $q$ in everything to follow)  is included, then
 the component of the vector potential $A_N$ normal to $\Sigma$ couples to
the normal part of the gradient. Given that $V_C$ is generated
from differentiations in $q$ and the requirement of conservation
of the norm, it is not surprising that other curvature effects
would follow from inclusion of $\mathbf A$. This work is concerned
with determining the effective potential that arises from the $A_N
{\partial / \partial q}$ operator, and through numerical
calculation with a simple but realistic model, gauging its
influence on the single particle spectra and wave functions as a
function of field strength and orientation.

The remainder of this paper is organized as follows: Section 2
develops the Schrodinger equation
\begin{equation}
{1 \over {2m}}\bigg ( {\hbar \over i} \nabla + e {\mathbf A}
\bigg) ^2\Psi = E\Psi
\end{equation}
for an electron allowed to move in the neighborhood $\Sigma$. In a
preprint version of this work \cite{enconeal} differential forms
were employed to represent Eq. (1), but here more conventional
language is used. Section 3 briefly reviews the procedure by which
$V_C$ is derived and uses a similar but not identical  methodology
to reduce the $A_N \partial /
\partial q$ operator to an effective potential expressed entirely
in surface variables. Section 4 employs the formalism presented in
section 3 to calculate spectra and wave functions as a function of
field strength and orientation for a toroidal structure, and
section 5 is reserved for conclusions.

\section{Curved surface Schrodinger equation}

The development of the Schrodinger equation in three dimensions on
an arbitrary manifold generally yields a cumbersome expression. To
remove some complexity in what ensues, an azimuthally symmetric
surface $\Sigma$ with $q$ the coordinate that gives the distance
from $\Sigma$ will be adopted.

 Let ${\mathbf
e}_{\rho},{\mathbf e}_{\phi},{\mathbf e}_{z}$ be cylindrical
coordinate system unit vectors. Parameterize $\Sigma(\rho,\phi)$
by the Monge form
\begin{equation}
{\mathbf r}(\rho,\phi)=\rho \ {\mathbf e}_{\rho} + S(\rho)\
{\mathbf e}_{z}
\end{equation}
with  $S(\rho)$ the shape of the surface. Points near the
 surface $S(\rho)$ are then described by
\begin{equation}
{\mathbf x}(\rho,\phi,q)={\mathbf r}(\rho,\phi)+q \ {\mathbf
e}_{n}(\rho,\phi)
\end{equation}
\noindent with ${\mathbf e}_{n}$ everywhere normal to the surface
and to be defined momentarily below. The differential line element
of Eq.(3) is
\begin{equation}
{d\mathbf x}=d{\mathbf r} + dq \ {\mathbf e}_{n}+ q \ d {\mathbf
e}_{n}.
\end{equation}
After some manipulation along with a list of expressions to be
defined below, Eq. (4) can be rewritten as (the subscript on $S$
denotes differentiation with respect to $\rho$)

\begin{equation}
{d\mathbf x}=Z(1+k_1q){\mathbf e}_1d\rho+ \rho(1+k_2q){\mathbf
e}_{\phi}d\phi+ dq   {\mathbf e}_{n}
\end{equation}
$$
\equiv Z f_1 {\mathbf e}_1d\rho+ \rho f_2 {\mathbf e}_{\phi}d\phi+
dq {\mathbf e}_{n}
$$
with
\begin{equation}
Z = \sqrt {1 + S^2_\rho}
\end{equation}
\begin{equation}
{\mathbf e}_1 = {1 \over Z}({\mathbf e}_\rho + S_\rho {\mathbf
e}_z)
\end{equation}
\begin{equation}
 {\mathbf e}_n = {1 \over Z}(-S_\rho{\mathbf e}_\rho +
{\mathbf e}_z)
\end{equation}
and the principle curvatures
\begin{equation}
k_1 =-{ S_{\rho\rho} \over Z^3},
\end{equation}
\begin{equation}
k_2 =-{ S_{\rho} \over \rho Z}.
\end{equation}

The metric elements can be read off of
\begin{equation}
dx^2=  Z^2 f_1^2 \ d\rho^2 + \rho^2 f_2^2\ d\phi^2 +\ dq^2
\end{equation}
from which  the Schrodinger equation can be determined,  but since
the minimal prescription will be employed as per Eq. (1) it proves
convenient to use the gradient
\begin{equation}
\nabla = {1\over f_1 Z}{\mathbf e}_1 {\partial \over \partial
\rho}+ {1\over f_2 \rho}{\mathbf e}_\phi {\partial \over \partial
\phi}+ {\mathbf e}_n {\partial \over \partial q}
\end{equation}
instead. Eq. (1) can be rearranged to
%\begin{widetext}
$$
{1 \over 2}\bigg[ {1\over Z^2 f_1^2} {\partial^2 \over \partial
\rho^2}+ {1 \over f_1 f_2 Z} {1 \over \rho}{\partial  \over
\partial \rho}+{1 \over f_1 f_2} {1 \over \rho^2}{\partial^2
\over
\partial \phi^2}+\bigg({k_1 \over f_1} + {k_1 \over f_1}
\bigg){\partial \over \partial q}+{\partial^2 \over \partial q^2}
$$
$$
  - \bigg({\rho
k_1 k_2 \over f_1^2} + {k_1 \over f_1} + {q k_{1\rho}\over Z^2
f_1^3} \bigg) {\partial \over \partial \rho}+ 2 \lambda i \bigg(
{A_1 \over f_1 Z}{\partial \over \partial \rho} + {A_\phi \over
f_2 \rho}{\partial \over \partial \phi} +{A_N }{\partial \over
\partial q}\bigg)
$$
\begin{equation}
 - \lambda^2\big(A_1^2 + A_\phi^2 + A_N^2 \big)+ {2Em \over \hbar^2} \bigg] \Psi =
 0
\end{equation}
with $\lambda = {e / \hbar}$ and $A_j = \bf A \cdot e_j$.
%\end{widetext}

While Eq. (13) describes the general case for an azimuthally
symmetric geometry, it can be simplified substantially when
considering the $q \rightarrow 0$ limit, or as dubbed by Golovnev
\cite{golovnev}, with thin layer quantization. The procedure
entails first performing all $q$ differentiations in accordance
with the intuitive notion that the kinetic energy in a thin layer
is large, then setting $q = 0$ everywhere. Eq. (13), leaving the
$q$-differentiations intact and setting $q=0$ everywhere save the
 $A_N$ term, cleans up to
\vskip 3pt
$$
{1 \over 2}\bigg[ {1\over Z^2 } {\partial^2 \over \partial
\rho^2}+ {1 \over Z} {1 \over \rho}{\partial  \over \partial
\rho}+ {1 \over \rho^2}{\partial^2  \over \partial \phi^2}+ 2h
{\partial \over
\partial q}+{\partial^2 \over \partial q^2}
$$
$$
  - k_1\big( \rho k_2 +1 \big)    {\partial \over
\partial \rho}+ 2 \lambda i \bigg( {A_1 \over Z}\bigg|_{q=0}{\partial \over
\partial \rho} + {A_\phi \over \rho}\bigg|_{q=0}{\partial \over \partial \phi}
+{A_N }{\partial \over
\partial q}\bigg)
$$
\begin{equation}
 - \lambda^2\big(A_1^2 + A_\phi^2 + A_N^2 \big)\bigg|_{q=0}+ {2Em \over \hbar^2} \bigg] \Psi =
 0
\end{equation}
with the mean curvature $h$ above given by
\begin{equation}
h={1\over 2}(k_1+k_2).
\end{equation}
The Gaussian curvature which will appear later is
\begin{equation}
k = k_1 k_2.
\end{equation}

\section{Derivation of geometric potentials}

The geometric potential $V_C(\rho)$ is found by reducing Eq. (13)
further by the well known procedure of assuming a suitable
confining potential in the normal direction $V_n(q)$ and demanding
conservation of the norm in the $q \rightarrow 0$ limit. The
latter requirement is generally imposed via  assuming a separation
of variables in the surface and normal parts of the total wave
function
\cite{burgsjens,jenskoppe,dacosta1,dacosta2,matsutani,goldjaffe,
exnerseba,schujaff}
\begin{equation} \Psi(\rho,\phi,q) \rightarrow
\chi_S(\rho,\phi)\chi_N(q)
\end{equation}
and imposing conservation of the norm through
\begin{equation}
|\Psi(\rho,\phi,q)|^2 (1+2qh+q^2k) d\Sigma dq \rightarrow
|\chi_S(\rho,\phi)|^2|\chi_N(q)|^2 d\Sigma dq
\end{equation}
or equivalently
\begin{equation}
\Psi =   \chi_S \chi_N (1+2qh+q^2k)^{-1/2} \equiv \chi_S \chi_N
G^{-1/2}.
\end{equation}

Inserting the rightmost term  in Eq. (19) into Eq. (13) and
subsequently taking $q\rightarrow 0$ reduces the $q$
differentiations there to
\begin{equation}
2h {\partial \over \partial q}+{\partial^2 \over \partial q^2}
\rightarrow {\partial^2 \over \partial q^2} + h^2-k.
\end{equation}
Separability of the surface and normal variables in the
non-magnetic part of the Hamiltonian is manifest.

Since $A_N$ can be a function of $q$ while $h$ is not, it is not
immediately apparent that separability of the Schrodinger equation
in the surface and normal variables is preserved by the
$A_N(\rho,\phi,q) {\partial / \partial q }$ operator. Rather than
apply the  identical procedure above to this term, instead
integrate out any $q$-dependence with some reasonable ansatz for
$\chi_N(q)$, say a normalized hard wall form $\chi_N(q) = {\sqrt
{2 / L}}\ {\rm sin} {\pi q /L}$. While equivalent to the method
summarized above in Eqs. (17)-(20),  it assists in establishing
conditions on $A_N$ that will prove useful later, and previous
work \cite{encscripta} has demonstrated that to a good
approximation this procedure is justified even if a particle is
not strongly confined to a region near $\Sigma$.

Write
\begin{equation}
I=\int_0^L {\chi_N(q) \over G^{-{1/ 2}}}A_N(\theta,\phi,q)
\bigg[{\partial \over
\partial q }{\chi_N(q)\over G^{-{1/2}}}\bigg]
G dq
\end{equation}
\vskip 8pt \noindent which is equivalent to
\begin{equation}
I=-\int_0^L \chi_N^2(q)(h+qk) G^{-1} A_N(\rho,\phi,q)dq + \int_0^L
\chi_N(q)A_N(\rho,\phi,q)  \chi'_N(q)dq.
\end{equation}
Consider the left hand integral $I_L$, and assume (suppressing
surface arguments) $A_N(q)\sim a_0+a_1q+...$; expanding out the
arguments and noting that each power of $q$ when integrated picks
up a power of $L$ that will vanish as $L \rightarrow 0$, the
result in this limit is
\begin{equation}
I_L^0\cong-hA_N(\rho,\phi,0)\int_0^L \chi_N^2(q)dq
\end{equation}
i.e., the effective potential arising from $I_L$ is proportional
to $-hA_N(\rho,\phi,0)$.

Turning to the second integral $I_R$, write $\chi_N(q)\chi'_N(q)$
as ${\partial /
\partial q}[{\chi^2_N(q)/2}]$ and perform an integration by parts.
The surface term vanishes  so that
\begin{equation}
 I_R = -{1 \over 2}\int_0^L \chi_N^2(q){\partial A_N(\rho,\phi,q) \over
\partial q } dq.
\end{equation}
Again taking $A_N(q)\sim a_0+a_1q+...$ allows Eq. (24) to be
expressed in the $q \rightarrow 0$ limit as
\begin{equation}
I_R^0 \cong -{1 \over 2}\int_0^L \chi_N^2(q){\partial
A_N(\rho,\phi,q) \over
\partial q }\bigg |_{q=0} dq.
\end{equation}
 Eqs. (23) and (25) combine to give an effective surface
potential (with constants appended)
\begin{equation}
V^{mag}_N(\rho,\phi)={i e \hbar \over m} \bigg[
h(\rho)A_N(\rho,\phi,0)+ {1 \over 2}{\partial A_N (\rho,\phi,q
)\over
\partial q } \bigg |_{q=0} \bigg ].
\end{equation}
There are two points that should be addressed with  further
explanation. First,  assuming a nonsingular series expansion for
$A_N$ is reasonable since $A_N(\rho,\phi,0)$ is the value of the
vector potential on the surface; it should certainly be physically
well behaved there but need not vanish. Secondly, while a hard
wall form has been assumed for the trial wave function (a Gaussian
works as well), it is a reasonable
 conjecture that the arguments made above are independent of the
choice of $\chi_N(q)$ as long as there is negligible mixing
amongst states in the $q$ degree of freedom.

\section{Numerics}
The formalism developed above is now applied to calculate the
spectrum and eigenfunctions for a toroidal structure in a uniform
magnetic field of arbitrary orientation inclusive of geometric
potentials.

A convenient choice to parameterize points near a
   toroidal surface $T^2$ of major radius $R$ and minor radius
 $a$ is \cite{fpl}
\begin{equation}
 \mathbf{x} (\theta,\phi)=W (\theta){\mathbf e}_\rho +a\  {\rm sin}
\theta{\mathbf e}_z  + q {\mathbf e}_n
\end{equation}
with
\begin{equation}
 W = R + a \ {\rm cos} \theta,
\end{equation}
and
\begin{equation}
{\mathbf e}_n = {\rm cos}\theta{\mathbf e}_\rho +  {\rm sin}\theta
{\mathbf e}_z.
\end{equation}
The differential line element is
\begin{equation}
 d \mathbf{x}=a(1+k^T_1q){\mathbf e}_\theta  d\theta
+W(\theta)(1+k^T_2q){\mathbf e}_\phi d\phi + {\mathbf e}_n dq
\end{equation}
$$
\equiv a_q(q){\mathbf e}_\theta  d\theta + W_q(\theta,q) {\mathbf
e}_\phi d\phi +  {\mathbf e}_n dq
$$
with ${\mathbf e}_\theta =-\rm sin \theta  {\mathbf e}_\rho +\rm
cos \theta {\mathbf e}_z$ and the toroidal principle curvatures
\begin{equation}
k_1^T = {1 \over a},
\end{equation}
\begin{equation}
k_2^T = {{\rm cos}\theta \over W(\theta)}.
\end{equation}
 The gradient that follows from Eq. (30)
is

\begin{equation}
\nabla = {\mathbf e}_\theta {1 \over a_q(q)} {\partial \over
\partial \theta}+ {\mathbf e}_\phi {1 \over W_q(\theta,q)} {\partial \over \partial
\phi}+ {\mathbf e}_n {\partial \over \partial q}.
\end{equation}

 The symmetry of the torus allows an arbitrary static magnetic
 field to be taken as
\begin{equation}
{\mathbf B} = B_1{\mathbf i} + B_0{\mathbf k}.
\end{equation}
 In the Coulomb gauge
 the vector potential ${\mathbf
A}(\theta,\phi) = {1 \over 2} \mathbf{B} \times \mathbf{r} $
expressed in the geometry of Eq. (27) is

\begin{equation}
\notag   \mathbf {A}(\theta,\phi,q) =    {1\over 2}\big [ B_1 {\rm
sin}\phi (R {\rm cos} \theta  + a_q){\mathbf e}_\theta  +
 (B_0 W_q - B_1 a_q \ {\rm   sin \theta cos\phi}){\mathbf e}_\phi
\end{equation}
\begin{equation} + B_1R
{\rm sin\phi \sin\theta} {\mathbf e}_n].
\end{equation}
In this case $\partial A_N / \partial q = 0$ so only the first
term in Eq. (26) will contribute to $V^{mag}_N(\theta,\phi)$. The
Schrodinger equation in the $q \rightarrow 0$ limit inclusive of
geometric potentials can be written (spin will be neglected) in a
compact form by first defining
$$\alpha = a/R$$
$$ F(\theta) = 1 + \rm \alpha \ cos\theta$$
$$  \gamma_0 = B_0 \pi R^2 $$
$$ \gamma_1 = B_1 \pi R^2 $$
$$ \gamma_N = {\pi  \hbar \over e} $$
$$ \tau_0 = {\gamma_0 \over \gamma_N}$$
$$ \tau_1 = {\gamma_1 \over \gamma_N}$$
$$ \varepsilon = -{2mEa^2 \over \hbar^2},$$
after which Eq. (1) may be written
$$
 \bigg [ {\partial^2 \over \partial \theta^2} -
  {\alpha \  {\rm sin} \ \theta \over F(\theta)}{\partial \over \partial
 \theta} + {\alpha^2 \over F^2(\theta)}{\partial^2 \over
\partial \phi^2} + {1\over 4 F^2(\theta)}+{i\alpha\tau_1 \over 2}
{\rm sin\theta sin\phi}{ {(1+2\alpha \ {\rm cos}\theta)} \over
F(\theta)}
$$
$$
+ i \bigg(\tau_0\alpha^2-{\tau_1\alpha^3 \over F(\theta)}{\rm
sin\theta cos\phi} \bigg){\partial \over \partial \phi}
 + i\alpha\tau_1 {\rm sin \phi (\alpha+cos\theta)}{\partial
\over
\partial \theta}
$$
\begin{align}
 -{\tau_0^2 \alpha^2F^2(\theta) \over 4} - {\tau_1^2 \alpha^2 F^2(\theta) \over 4}
\bigg ({\rm sin^2}\phi + {\alpha^2 \ {\rm sin^2}\theta \over
F^2(\theta)}\bigg) +{\tau_0 \tau_1 \alpha^3 F(\theta) \over 2}\rm
sin\theta cos\phi
  \bigg] \Psi = \varepsilon\Psi
\end{align}
\begin{equation}
\Rightarrow H_\tau\ \Psi = \varepsilon \Psi,
\end{equation}
with the fourth and fifth terms of Eq. (36) being proportional to
$V_C$ and $V^{mag}_C$.

To obtain solutions of Eq.(36) a basis set expansion
 orthogonal over the integration
measure $dJ(\theta)= F(\theta)d\theta d\phi$  may be employed for
a given $\alpha$.  Here $R$ will be set to $500 \AA$ in accordance
with fabricated structures \cite{garsia, lorke, lorke2,zhang} (for
an $R =500 \AA$ torus $\tau = .263 B_0$).  With $a = 250 \AA$,
$\alpha$ = $1/2$, a value which serves as a compromise between
smaller $\alpha$ where the solutions tend towards simple
trigonometric functions and larger $\alpha$ which are less likely
to be physically realistic.

There are two options available for basis functions, one being
\begin{equation}
\chi_{n\nu}(\theta,\phi) = {1 \over \sqrt {2 \pi}}F^{-{1 \over 2}}
e^{i n \theta}e^{i\nu\phi}
\end{equation}
and the other  being
\begin{equation}
\chi^{\pm}_{n\nu}(\theta,\phi) =  \left ( \begin{array}{c} f_n(\theta) \\
g_n(\theta) \end{array} \right)e^{i\nu\phi}.
\end{equation}
with $f_n(\theta),g_n(\theta)$ even/odd  orthonormal functions
labelled by +/- respectively,  constructed by a Gram-Schmidt (GS)
procedure \cite{gst2} over $dJ(\theta)$ using ${\rm cos} \
n\theta, {\rm sin}\ n \theta$ as primitives. Here the latter
approach will be adopted in order to facilitate comparisons to a
related work \cite{encT2mag} and because of the ease by which the
Hamiltonian matrix elements

\begin{equation}
H_{ \bar{n} \bar{\nu} n \nu}^{\bar{\pi}\pi}= \big
<\chi^{\bar{\pi}}_{\bar{n} \bar{\nu}} |H_\tau| \chi^{\pi}_{n \nu}
\big >
\end{equation}
can be evaluated analytically. The basis comprises
 six GS functions of each $\theta$-parity and five
azimuthal functions spanning $-2 \leq \nu \leq 2$ per
$\theta$-function for a total of 60 basis states. The resulting 60
x 60 Hamiltonian matrix blocks schematically into

\[  \left( \begin{array} {cc}
  H^{++}& H^{+-} \\
  H^{-+} & H^{--} \end{array} \right)\]
 from which eigenvalues and eigenfunctions are determined. Since
 the concern here is with the ground state eigenfunctions only a
 few GS states prove relevant; they are $\ f_0(\theta)  = .3987$,
$\ f_1(\theta) = .6031 \ \rm cos \theta-.1508$ and $\ g_1(\theta)
= .5642 \rm \ sin \theta$.

Figs. (1-3) plot the ground state energy $\varepsilon_0$ as a
function of the magnitude of flux $\tau$ for three field
orientations.  Each figure displays results for
$\varepsilon_0(\tau)$ with $V_C$ and/or $V^{mag}$ switched on or
off. In Fig. 1 the field is oriented along the z-axis. In this
case there there is no component of $\mathbf A$ normal to the
surface so only two curves are evidenced. The curves are
qualitatively similar and the effect of $V_C$ is to smooth out the
$V_C = 0$ curve at $\tau_0 \approx 1$ and shift it downward by an
overall constant. It should be noted that although the curves are
similar, persistent current effects depend of the smoothness and
shape of the of $\varepsilon(\tau)$ curves so that even fine
details can prove important. Table I shows the evolution of the
ground state wave function $\chi_G(\theta,\phi)$ for  several
$\tau$; because the field is oriented along the z-axis, $\tau$
also measures the flux through the toroidal plane.  Both $V_C$ and
$V^{mag}$ are zero so that the evolution of $\chi_G(\theta,\phi)$
is  due only  to changes in field strength.

 Fig. 2 shows results for a field orientation oriented tilted $\pi \over
4$ radians relative to the toroidal plane with the magnitude of
$\tau$  plotted on the horizontal axis. The divergence of the two
lower curves illustrates the influence of $V^{mag}$ on the
spectrum. The curve inclusive of $V^{mag}$ begins to involve
excited states in both the $\theta$ and azimuthal degrees of
freedom (see table II). While the admixture of excited states
would tend to raise $\varepsilon_0$, the interaction is strong
enough to pull the $\varepsilon_0(\tau)$ down as the applied field
increases.

In Fig. 3 the magnetic field is situated parallel to the toroidal
plane along the x-axis. As would be anticipated, there is no
structure in the $V_C = V^{mag} =0 $ curve since no flux
penetrates the plane of the torus. This trend obtains also when
$V_C,V^{mag}\neq 0$. The effect of $V^{mag}$ becomes substantial
very quickly both in $\varepsilon_0(\tau)$ and on
$\chi_G(\theta,\phi)$ (table III). In this case there is not even
qualitative agreement between results with $V_C \neq 0$ with
$V^{mag}$   omitted  compared to  when it is included.

\section{Conclusions}

This work presents a method to reduce the
  $A_N \partial / \partial q$ term appearing in the
Schrodinger equation for an electron near a two-dimensional
surface in an arbitrary static magnetic field to a geometric
potential $V^{mag}$ written entirely in terms of surface
variables. This potential can appreciably modify energy vs.
magnetic flux curves as well as surface wave functions considered
here.

In the context of real structures, the practical utility of
deriving geometric potentials lies in reducing  three dimensional
problems to two-dimensional ones. Earlier work \cite{encscripta}
on a simple nanoscale model has shown that solutions of an
ab-initio two dimensional Schrodinger equation do not adequately
approximate thin layer three-dimensional solutions on a curved
space unless $V_C$ is included in the Schrodinger equation. From
this perspective, geometric potentials of the form discussed here
should  be considered effective potentials that must be included
in  the modelling of curved structures in order to achieve a
complete description of the object.

\begin{center} {\bf ACKNOWLEDGEMENTS} \end{center} The author would like to thank M.
Jack for useful discussions.
\newpage

\bibliography{refcomp}

\newpage
%TABLE 1
\begin{table}
\caption{Ground state wave functions for ${\mathbf B} =
B_0{\mathbf k}$ at integer values of $\tau$. The arguments in
square brackets indicate if $V_C$/ $V^{mag}$ are switched off or
on. Only the dominant terms are shown. }
\begin{center}
\begin{tabular}{|c|c|c|c|}
\hline
 & $\tau = 0$ & $\tau = 1$ & $\tau = 2$ \\
\hline
$\chi_G$[off,off] & 1 & $\sim 1$ & $ (- .969 f_0+ .245f_1)e^{-i\phi}$\\
$\chi_G$[on,off]  & $ - .968 f_0- .244 f_1$ & $ - .957 f_0+ .254 f_1$ & $ ( .987 f_0 - .158 f_1)e^{-i\phi}$\\
$\chi_G$[on,on]  & $ - .968 f_0- .244 f_1$ & $ - .957 f_0+ .254 f_1$ & $ ( .987 f_0 - .158 f_1)e^{-i\phi}$\\
 \hline
\end{tabular}
\end{center}
\end{table}

%TABLE 2
\begin{table}
\caption{Ground state wave functions for ${\mathbf B} =
B_0({\mathbf i}+ {\mathbf k})/ \sqrt 2 $ as per table I. For this
field configuration $A_N(\theta,\phi) \neq 0$. }
\begin{center}
\begin{tabular}{|c|c|c|c|}
%\begin{tabular}{cp{5cm}}
\hline
 & $\tau = 0$ & $\tau = 1$ & $\tau = 2$ \\

\hline
$\chi_G$[off,off] & 1 & $-.989f_0 -.115g_1 e^{-i\phi}$ &
$(- .928 f_0+ .159f_1)e^{-i\phi}-$ \\ &  &  & $g_1(.287-.145e^{-i\phi})$\\

$\chi_G$[on,off] & $.968f_0-.244f_1$ & $.961f_0-.252f_1$ & $ .932f_0-.270f_1$\\

$\chi_G$[on,on] & $.968f_0-.244f_1 $ & $(.957 f_0- .232f_1)$+& $f_0(.909 e^{-i\phi}- .126)-$ \\
& & \ $  g_1(.094e^{i\phi}-.127e^{-i\phi})$ &
$g_1(.173e^{-i\phi}-.351)$ \\

 \hline
\end{tabular}
\end{center}
\end{table}

%TABLE 3
\begin{table}
\caption{Ground state wave functions for ${\mathbf B} =
B_1{\mathbf i}$ at integer values of $\tau$ as per table I.
$A_N(\theta,\phi)$ is nonzero, and there is no flux through
 the toroidal plane.}
\begin{center}
\begin{tabular}{|c|c|c|c|}
\hline
 & $\tau = 0$ & $\tau = 1$ & $\tau = 2$ \\
\hline
$\chi_G$[off,off] & $.968f_0-.244f_1$ & $.978 f_0 + .279 i g_1{\rm sin}\phi$ & $ .894 f_0 + .133f_1- .552 i g_1{\rm sin}\phi$\\
$\chi_G$[on,off]  & $  .968 f_0- .244 f_1$ & $  .964 f_0- .218 f_1$ & $ .941 f_0 - .132f_1+ .403i g_1{\rm sin}\phi$\\
$\chi_G$[on,on]  & $ .968 f_0- .244 f_1$ & $  -.954 f_0+ .178 f_1+.320ig_1{\rm sin}\phi $
& $ .869 f_0 - .250f_1{\rm cos \phi}- .314i g_1{\rm sin}\phi$\\
 \hline
\end{tabular}
\end{center}
\end{table}

\newpage

\begin{figure}
\centering
\includegraphics{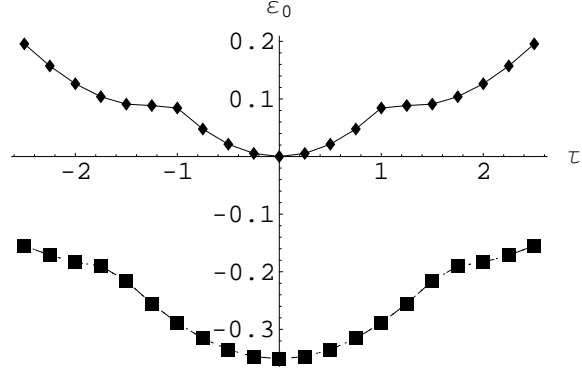}
\caption{$\varepsilon_0$ as a function of $\tau$ for $\mathbf B$ =
$B_0 \mathbf k$. Diamonds correspond to $V_C = V^{mag}=0$, stars
to $V_C \neq 0$, $ V^{mag}=0$ and squares to $V_C \neq 0$, $
V^{mag} \neq 0$. }
%\centerline{Fig. 1}
\end{figure}

\begin{figure}
\centering
\includegraphics{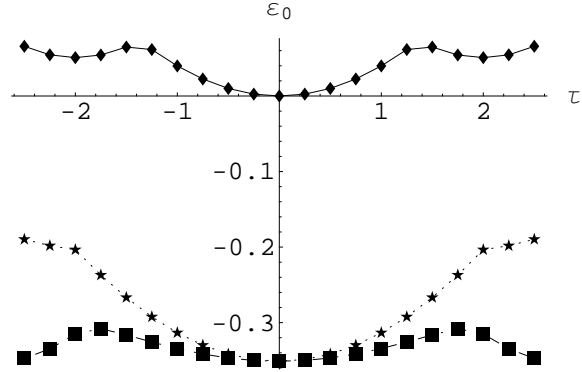}
\caption{$\varepsilon_0(\tau)$ for ${\mathbf B} = B_0({\mathbf i}+
{\mathbf k})/ \sqrt 2 $. Diamonds correspond to $V_C = V^{mag}=0$,
stars to $V_C \neq 0$, $ V^{mag}=0$ and squares to $V_C \neq 0$, $
V^{mag}
\neq 0$. }% \centerline{Fig. 2}
\end{figure}

\begin{figure}
\centering
\includegraphics{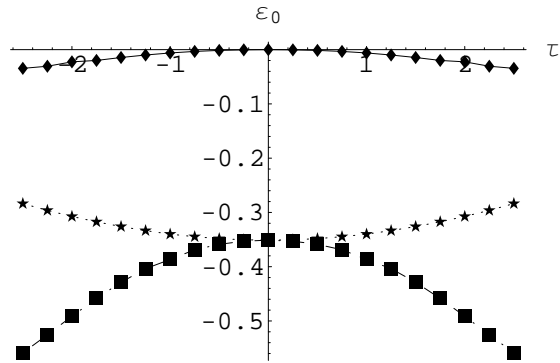}
\caption{$\varepsilon_0(\tau)$ for $\mathbf B$ = $B_1\mathbf i$.
Diamonds correspond to $V_C = V^{mag}=0$, stars to $V_C \neq 0$, $
V^{mag}=0$ and squares to $V_C \neq 0$, $ V^{mag}
\neq 0$. }% \centerline{Fig. 3}
\end{figure}

\end{document}